\documentclass[10pt,aps,prl,twocolumn,superscriptaddress]{revtex4-1}

\usepackage{amsmath}
\usepackage{textcomp}
\usepackage{amssymb} 
\usepackage{graphicx}
\usepackage{tabularx}
\usepackage{placeins}
\usepackage{natbib}
\usepackage[normalem]{ulem}
\usepackage[usenames]{color}

\makeatletter
\newcommand*{\rom}[1]{\expandafter\@slowromancap\romannumeral #1@} %roman numbers
\makeatother

\def\be{\begin{equation}}
	\def\ee{\end{equation}}
\def\bea{\begin{eqnarray}}
	\def\eea{\end{eqnarray}}
\def\ba{\begin{array}}
	\def\ea{\end{array}}

\begin{document}

\title{Mechanics of fiber networks under a bulk strain}

\author{Sadjad Arzash}
\altaffiliation[Present address: ]{Department of Physics, Syracuse University, Syracuse, NY}
\altaffiliation{Department of Physics \& Astronomy, University of Pennsylvania, Philadelphia, PA}
\affiliation{Department of Chemical \& Biomolecular Engineering, Rice University, Houston, TX 77005}
\affiliation{Center for Theoretical Biological Physics, Rice University, Houston, TX 77030}
\author{Abhinav Sharma}
\affiliation{Leibniz-Institut f\"ur Polymerforschung Dresden, Institut Theorie der Polymere, 01069 Dresden, Germany}
\author{Fred C.\ MacKintosh}
\affiliation{Department of Chemical \& Biomolecular Engineering, Rice University, Houston, TX 77005}
\affiliation{Center for Theoretical Biological Physics, Rice University, Houston, TX 77030}
\affiliation{Department of Chemistry, Rice University, Houston, TX 77005}
\affiliation{Department of Physics \& Astronomy, Rice University, Houston, TX 77005}

\begin{abstract}

Biopolymer networks are common in biological systems from the cytoskeleton of individual cells to collagen in the extracellular matrix. The mechanics of these systems under applied strain can be explained in some cases by a phase transition from soft to rigid states. For collagen networks, it has been shown that this transition is critical in nature and it is predicted to exhibit diverging fluctuations near a critical strain that depends on the network's connectivity and structure. Whereas prior work focused mostly on shear deformation that is more accessible experimentally, here we study the mechanics of such networks under an applied bulk or isotropic extension. We confirm that the bulk modulus of subisostatic fiber networks exhibits similar critical behavior as a function of bulk strain. We find different non-mean-field exponents for bulk as opposed to shear. We also confirm a similar hyperscaling relation to what previously found for shear.

\end{abstract}

\maketitle

\textit{Introduction} \textemdash\, The mechanical stability of cells and tissues is largely governed by interconnected biopolymer networks such as actin and collagen. In the linear regime, the rigidity of networks of stiff fibers such as collagen is strongly dependent on the average connectivity or coordination $z$ \cite{wyart_elasticity_2008,broedersz_criticality_2011,das_redundancy_2012}. While these structures undergo a mechanical phase transition from a floppy to a rigid state at the critical or \textit{isostatic} connectivity $z_c$ \cite{maxwell_i.reciprocal_1870}, this threshold lies well above the physiological connectivity of fibrous networks in 3D \cite{lindstrom_biopolymer_2010,lindstrom_finite-strain_2013,jansen_role_2018}. Such subisostatic fiber networks, however, undergo a strain-controlled phase transition from floppy to rigid states when subject to a finite deformation \cite{sheinman_nonlinear_2012,sharma_strain-controlled_2016}. This transition is critical and occurs at a threshold shear strain that depends on the network's connectivity and geometry \cite{feng_nonlinear_2016,sharma_strain-driven_2016,licup_elastic_2016}. More recent work has shown that the mechanical stability of these networks can be understood in terms of emerging states of self-stress \cite{vermeulen_geometry_2017,mao_maxwell_2018,rens_rigidity_2019,damavandi_energetic_2022,damavandi_energetic_2022-1}. Near the critical strain, the shear modulus exhibits a power law behavior with a non-mean-field exponent.

\begin{figure}[!h]
	\centering
	\includegraphics[width=7cm,height=7cm,keepaspectratio]{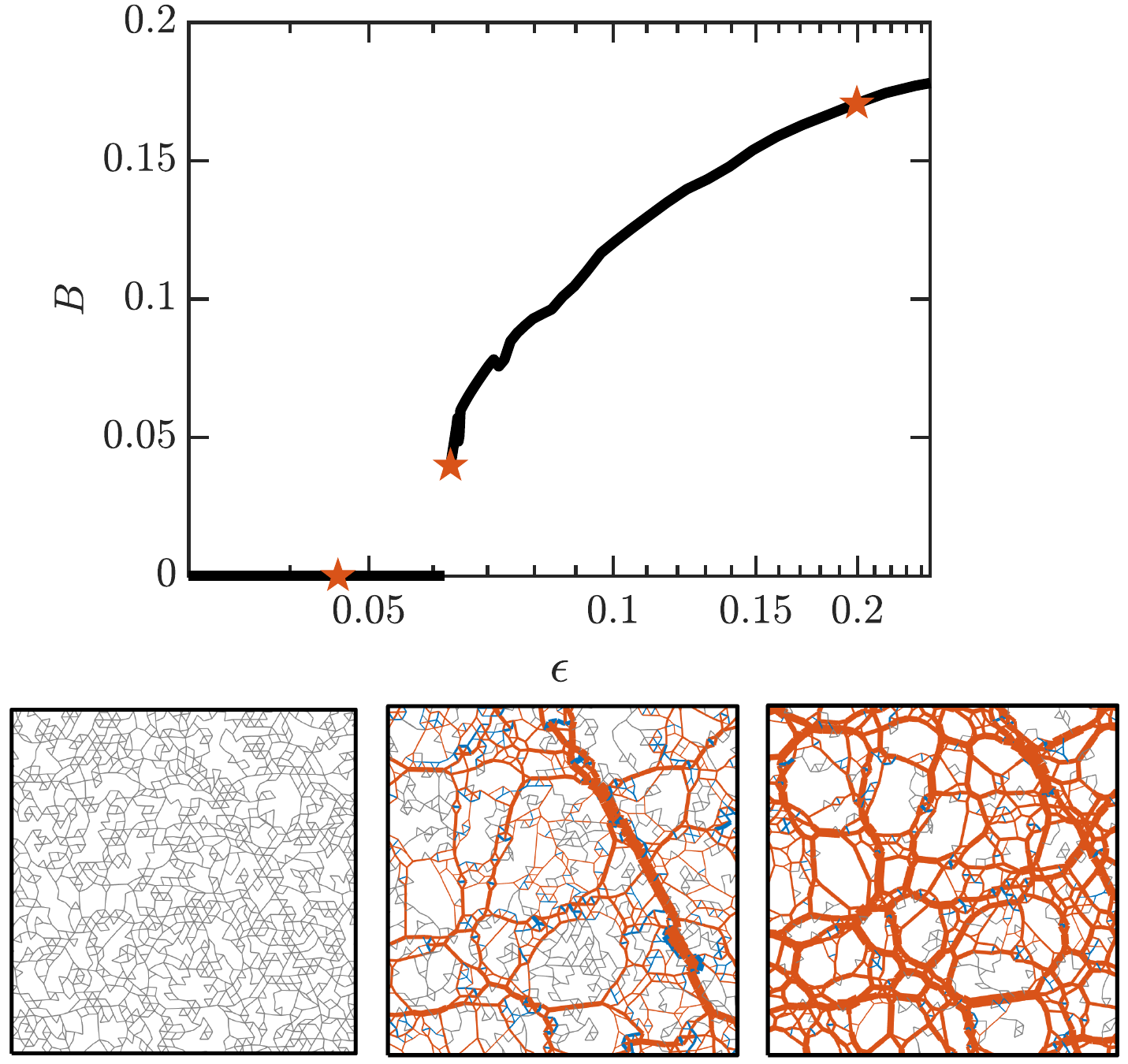}
	\caption{\label{fig_network} Nonlinear bulk modulus $B$ as a function of bulk strain for a diluted triangular network at $z=3.3$. The network has a zero bulk modulus below a critical strain at which $B$ jumps to a finite value $B_c$. Three snapshots of a small section of the network corresponding to the highlighted points on the modulus curve are shown. The frame is fixed in place for all three snapshots. For a bond $m$ with a tension $\tau_m = (l_m - l_{m,0})/l_{m,0}$, we plot its thickness proportional to its relative tension $\tau_m/\langle |\tau| \rangle$. The red and blue colors correspond to a positive and a negative $\tau_m$, respectively. The gray bonds have zero tension.}
\end{figure}

Here, we study the mechanics of subisostatic fiber networks under a finite isotropic extension. Using a 2D triangular lattice-based model, we calculate the nonlinear bulk modulus $B$ with varying bending stiffness of fibers. In the absence of bending interactions between fiber crosslinks, the bulk modulus remains zero until a critical extensional strain $\varepsilon_c$ at which $B$ jumps to a finite value (Fig.\ \ref{fig_network}). The onset of a finite bulk modulus coincides with a network-spanning tensional pattern of bonds, as shown in Fig.\ \ref{fig_network}. Near the critical extensional strain, we obtain various critical exponents and verify collapse of our modulus data using a Widom-like scaling function with two branches for strains above and below the threshold value. Moreover, we confirm a hyperscaling relation analogous to what has been previously derived for fiber networks under a simple shear strain \cite{shivers_scaling_2019}.

%\FloatBarrier

\textit{Model} \textemdash\, In order to study the mechanics of fiber networks under bulk extension, we begin with a triangular lattice in 2D. For every node or crosslink in a triangular structure, we have three well-defined crossing fibers. Thus, a full triangular network has a connectivity of 6. To avoid the trivial effect of a system-spanning fiber, we initially cut a random bond from each fiber \cite{das_effective_2007,broedersz_criticality_2011}. Consequently in order to mimic realistic subisostatic biopolymer networks, we reduce this connectivity to $z=3.3 < z_c=2d$ by randomly removing bonds from the initial full lattice. We note that the crosslinks in our model are permanent and freely hinging. Network's elastic energy include both stretching and bending terms
\begin{equation}\label{eq_energy}
	\mathcal{H} = \frac{\mu}{2} \sum_{ \langle ij \rangle}^{}\frac{ (l_{ij} - l_{ij,0}) ^2 }{l_{ij,0}} + \frac{\kappa}{2} \sum_{ \langle ijk \rangle}^{} {\frac{ (\theta_{ijk} - \theta_{ijk,0} )^2}{ l_{ijk,0} } },
\end{equation}
where $l_{ij,0}$ and $l_{ij}$ are the initial (relaxed) and current bond length between nodes $i$ and $j$, $\theta_{ijk,0}$ and $\theta_{ijk}$ are the initial and current angle between neighboring bonds $ij$ and $jk$, respectively, and $l_{ijk,0} = (l_{ij,0}+l_{jk,0})/2$. Here, $\mu$ is the stretching modulus and $\kappa$ is the bending rigidity of the fibers. The first summation is over all connected nodes and the second is over all nearest-neighbor bonds on the same fiber, i.e., collinear adjacent bonds in the initial configuration. We set $\mu = 1$ in our simulations and vary the dimensionless bending stiffness $\tilde{\kappa} = \kappa/\mu\ell_0^2$, where $\ell_0=1$ is the average bond length in the undeformed lattice \cite{licup_stress_2015}.

Here, we study the nonlinear bulk modulus of diluted triangular networks under the following bulk deformation tensor
\begin{equation} \label{eq_bulk_tensor}
	\Lambda_{B}(\varepsilon) = \begin{bmatrix}
		1 + \varepsilon & 0 \\
		0 & 1 + \varepsilon
	\end{bmatrix},
\end{equation}
where $\varepsilon$ is the extensional strain. By applying a bulk strain $\varepsilon$, the network's volume changes as $V = V_0 (1 + \varepsilon)^2$. We deform the network in a step-wise manner in a periodic box and minimize the elastic energy in Eq.\ \eqref{eq_energy} using FIRE \cite{bitzek_structural_2006} at each strain point to obtain the minimum energy configuration. The stress components are calculated by
\begin{equation} \label{eq_virial_stress}
	\sigma_{\alpha \beta} = \frac{1}{2V} \sum_{ij}^{}f_{ij,\alpha} r_{ij,\beta}
\end{equation}
where $V$ is the volume (area) of system, $f_{ij,\alpha}$ is the $\alpha$ component of the force exerted on node $i$ by node $j$, and $r_{ij,\beta}$ is the $\beta$ component of the displacement vector connecting nodes $i$ and $j$. The summation is taken over all nodes in the network. We find the nonlinear bulk modulus $B$ as
\begin{equation} \label{eq_bulk_modulus}
	B = - V \frac{\partial P}{\partial V} = V \frac{\partial \sigma_{\perp}}{\partial V} = \frac{1}{2} (1 + \varepsilon) \frac{\partial \sigma_{\perp}}{\partial \varepsilon},
\end{equation}
where the pressure of the system is $P = - \sigma_{\perp} = - \frac{1}{d}(\sum_{i}\sigma_{ii})$ in $d$ dimensions. The volume of the system is not preserved under an applied isotropic extension, i.e., $V = V_0 (1+\varepsilon)^d$ where $V_0$ is the initial volume. Here, we define the nonlinear bulk modulus in the deformed state of the network, which is different from a prior definition of this parameter \cite{sheinman_nonlinear_2012} that is defined in the undeformed volume. We note that we find the critical extensional strain $\varepsilon_c$ for every individual random sample using a bisection method. Unless otherwise stated, the results are averaged over 40 random samples. The error bars are the standard deviation for all samples.

\begin{figure}[!h]
	\centering
	\includegraphics[width=15cm,height=15cm,keepaspectratio]{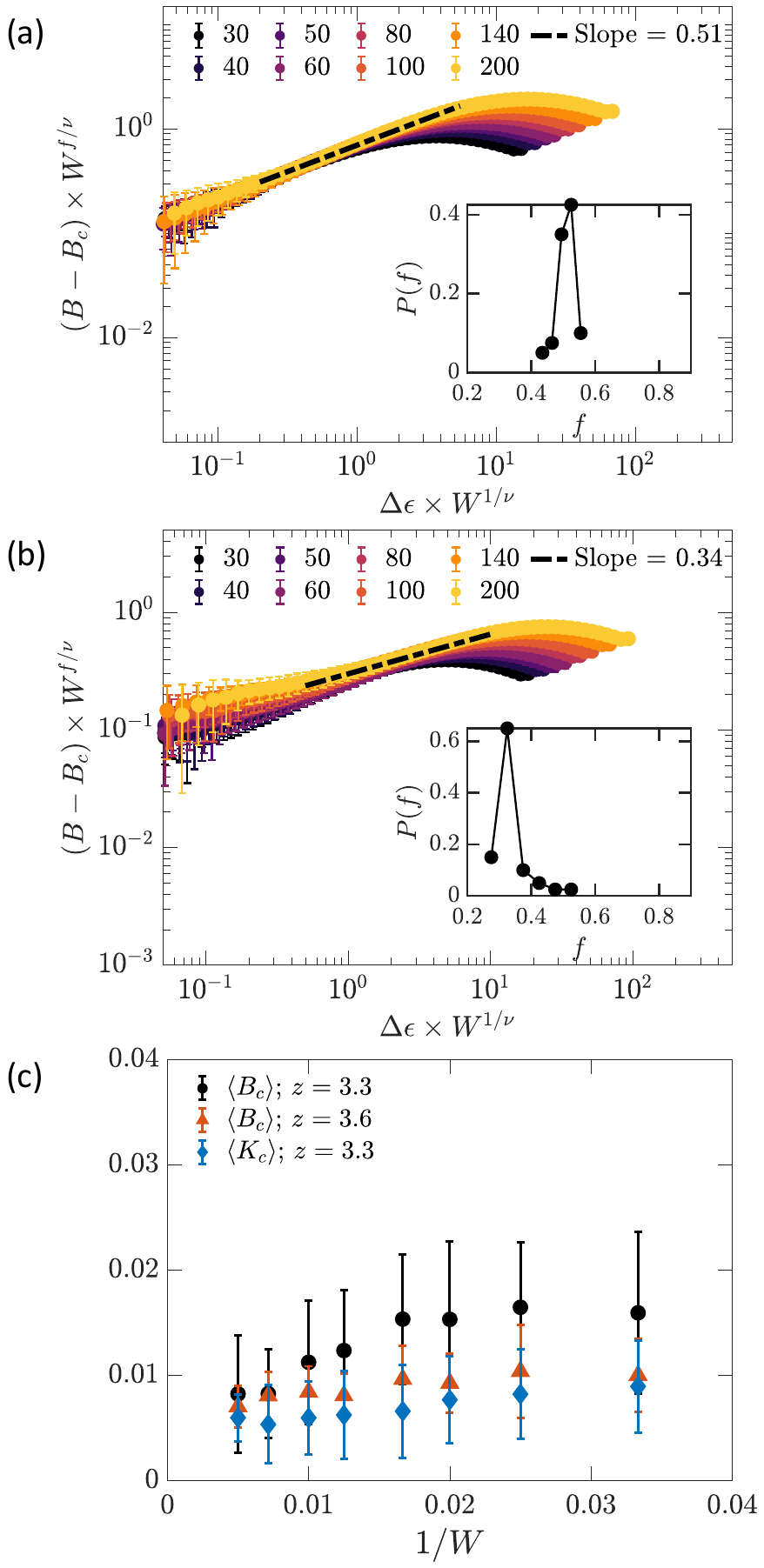}
	\caption{\label{fig_scaling_bulk_f} (a) Finite-size scaling analysis corresponding to Eq.\ \eqref{eq_scaling_bulk_f} for diluted triangular networks at $z=3.6$ with central-force interactions. In the critical region, we obtain $f = 0.51 \pm 0.03$. The distribution of $f$ is shown as an inset. (b) Similar finite-size scaling analysis of $B$ as in (a) but at a different connectivity $z=3.3$. In the critical region, we obtain $f = 0.34 \pm 0.05$. The distribution of $f$ is shown as an inset. (c) The ensemble average of the bulk modulus discontinuity $B_c$ for over 40 random samples of diluted triangular networks at $z=3.3$ and $z=3.6$ plotted versus inverse system size. The blue symbols, taken from Ref.\ \cite{arzash_finite_2020}, show the shear modulus discontinuity $K_c$ for the same system under a simple shear. To compare different values of $z$, we normalized these $B_c$ and $K_c$ values with the line density of networks in the undeformed state \cite{LineDensity}.}
\end{figure}
%\FloatBarrier

\textit{Results} \textemdash\, Similar to the behavior observed under an applied shear strain \cite{sharma_strain-controlled_2016,shivers_scaling_2019}, we find that subisostatic central-force networks undergo a floppy to a rigid state at a critical extensional strain $\varepsilon_c$ that depends on the connectivity and geometry of the system. At $\varepsilon_c$, the nonlinear bulk modulus exhibits a discontinuity $B_c$ in agreement with findings of Refs. \cite{merkel_minimal-length_2019,lee_stiffening_2022}, analogous to the behavior of the nonlinear shear modulus at the critical shear strain $\gamma_c$ \cite{vermeulen_geometry_2017,merkel_minimal-length_2019,rens_theory_2019,arzash_finite_2020}.

As shown in Fig.\ \ref{fig_scaling_bulk_f}c, the bulk modulus discontinuity at $\varepsilon_c$ appears to be larger than the shear modulus discontinuity at $\gamma_c$, both with a decreasing trend as we increase the system size. In the thermodynamic limit, we observe both finite $B_c$ and $K_c$. The jump in nonlinear bulk modulus at $\varepsilon_c$ is reminiscent of the behavior of linear bulk modulus of particulate systems at jamming \cite{goodrich_finite-size_2012,goodrich_jamming_2014} and generic Penrose tilings \cite{stenull_penrose_2014}. However, for the present strain-controlled transition, this discontinuity is not a reflection of a first-order transition. Rather, since strain is a control variable analogous to temperature, this discontinuity is more analogous to a discontinuity of the heat capacity in a thermal transition at a critical point \cite{shivers_scaling_2019}.

Near $\varepsilon_c$, we expect the following finite-size scaling relation to capture the behavior of $B$ in subisostatic central-force networks \cite{arzash_finite_2020}
\begin{equation} \label{eq_scaling_bulk_f}
	B - B_c = W^{-f/\nu} \mathcal{F}(\Delta \varepsilon W^{1/\nu}),
\end{equation}
where $\Delta \varepsilon = \varepsilon - \varepsilon_c$, $f$ is the bulk modulus scaling exponent in the regime $\varepsilon>\varepsilon_c$, $\nu$ is the correlation length exponent, and $\mathcal{F}(x)$ is a scaling function that is expected to increase as $x^{f}$ for large arguments, in order to obtain a well-defined thermodynamic limit. Although we use the same notation for the scaling exponents as for networks under shear, we note that the corresponding exponents may have different values. Figure\ \ref{fig_scaling_bulk_f}b shows the finite-size scaling behavior of $B$ in diluted triangular networks. In the critical region, we obtain $f = 0.34 \pm 0.05$ that is significantly smaller than the corresponding exponent $f = 0.79 \pm 0.07$ for the shear modulus. By increasing the connectivity of the network to $z=3.6$, we find that $f$ increases to $0.51 \pm 0.03$ (Fig.\ \ref{fig_scaling_bulk_f}a). We note that the dependence of critical exponents on $z$ for this transition has also been observed under a shear deformation \cite{sharma_strain-controlled_2016,arzash_shear-induced_2021}. By plotting the bulk modulus discontinuity at $\varepsilon_c$, we also find a decrease in $B_c$ with increasing system size (Fig.\ \ref{fig_scaling_bulk_f}c)), although with an apparent finite value in the thermodynamic limit.

\begin{figure}[!h]
	\centering
	\includegraphics[width=10cm,height=10cm,keepaspectratio]{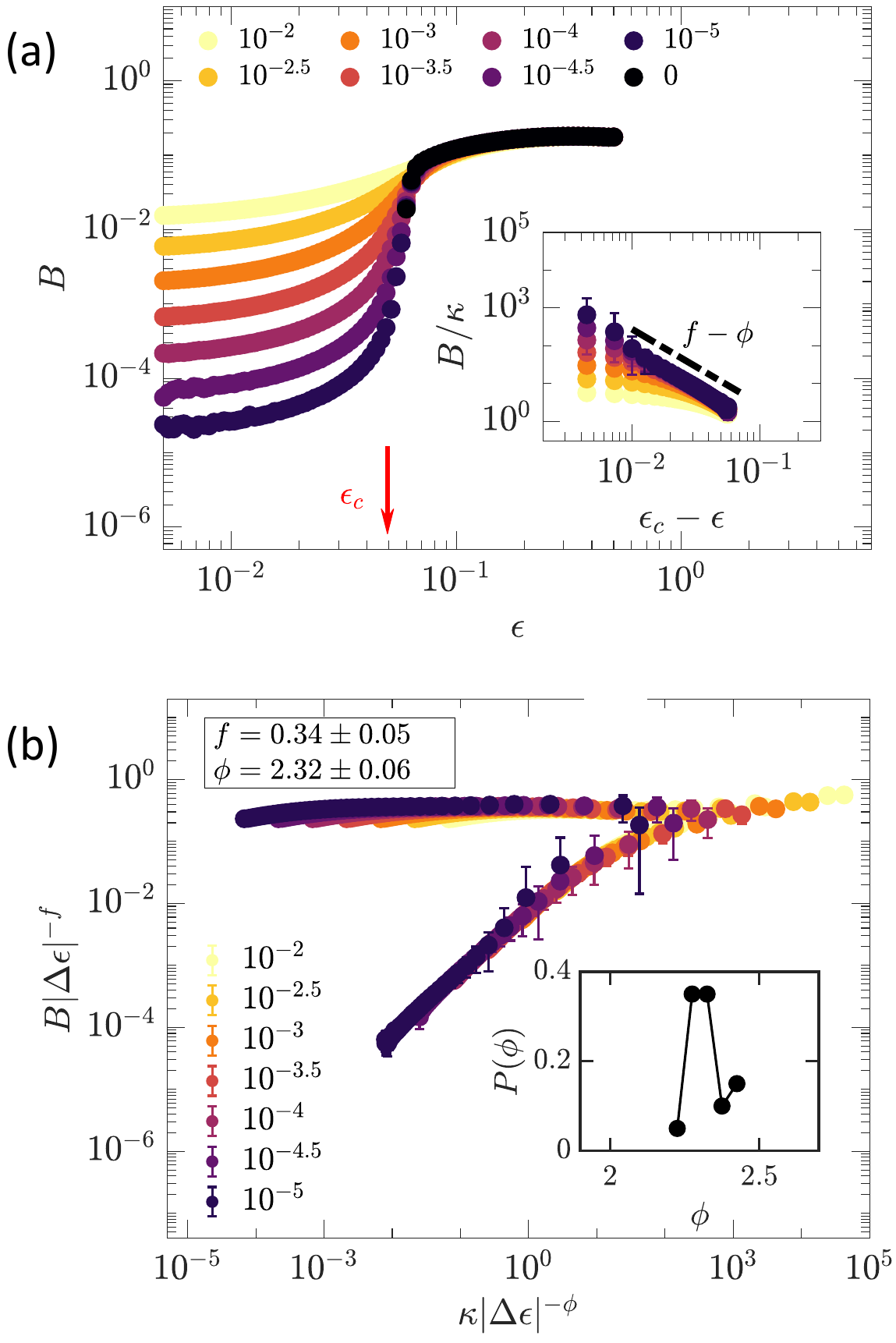}
	\caption{\label{fig_scaling_bulk_widom} (a) The nonlinear bulk modulus versus extensional strain for diluted triangular networks at $z=3.3$ and system size $W=100$ with varying bending rigidity $\kappa$ indicated in the legend. The data are averaged over 20 random samples. The inset shows the scaling behavior of $B$ in the subcritical region. Using the average value of the exponent $f$ obtained in Fig.\ \ref{fig_scaling_bulk_f}b, we find the exponent $\phi$ for individual samples by fitting a power law to the modulus data at $\kappa = 10^{-5}$. (b) The Widom-like scaling collapse of the modulus data in (a) using the average values of the exponents $f$ and $\phi$. The distribution of the exponent $\phi$ calculated from fitting a power law to the modulus data at $\kappa = 10^{-5}$ is shown as an inset.}
\end{figure}

By introducing bending interactions, the subisostatic networks become stabilized in the small strain regime $\varepsilon < \varepsilon_c$ with a bulk modulus proportional to the bending stiffness $B \sim \kappa$. This also has the effect of moving the system away from the critical singularity and suppressing the discontinuity in $B$. The nonlinear bulk modulus follows a Widom-like scaling relation given by
\be
B\approx\left|\varepsilon-\varepsilon_c\right|^f\mathcal{G}_{\pm}\left({\kappa}/{\left|\varepsilon-\varepsilon_c\right|^\phi}\right)\label{eq_scaling_bulk_widom}
\ee
for $\kappa > 0$, in which the branches of the scaling function $\mathcal{G}_{\pm}$ account for the bulk strain regimes above and below $\varepsilon_c$. Above the critical strain, $\mathcal{G}_{+}(x)$ is approximately constant for $x \ll 1$, while below the critical strain, we expect $\mathcal{G}_{-}(x) \sim x$ for $x \ll 1$, so that $B \sim \kappa |\varepsilon - \varepsilon_c|^{f-\phi}$. Near $\varepsilon_c$, however, continuity of $B$ requires $\mathcal{G}_{\pm}(x) \sim x^{f/\phi}$ for $x \gg 1$. Figure\ \ref{fig_scaling_bulk_widom}a shows $B$ versus $\varepsilon$ for triangular networks at $z=3.3$ with varying bending stiffness. In the subcritical region, we find $\phi = 2.32 \pm 0.06$. The data collapse of $B$ corresponding to Eq.\ \eqref{eq_scaling_bulk_widom} is shown in Fig.\ \ref{fig_scaling_bulk_widom}b. For networks with $z=3.6$, we obtain a Widom-like collapse of the modulus (not shown) similar to Fig.\ \ref{eq_scaling_bulk_widom}b using $\phi = 2.33 \pm 0.1$.

%\FloatBarrier

\begin{figure}[!h]
	\centering
	\includegraphics[width=7cm,height=7cm,keepaspectratio]{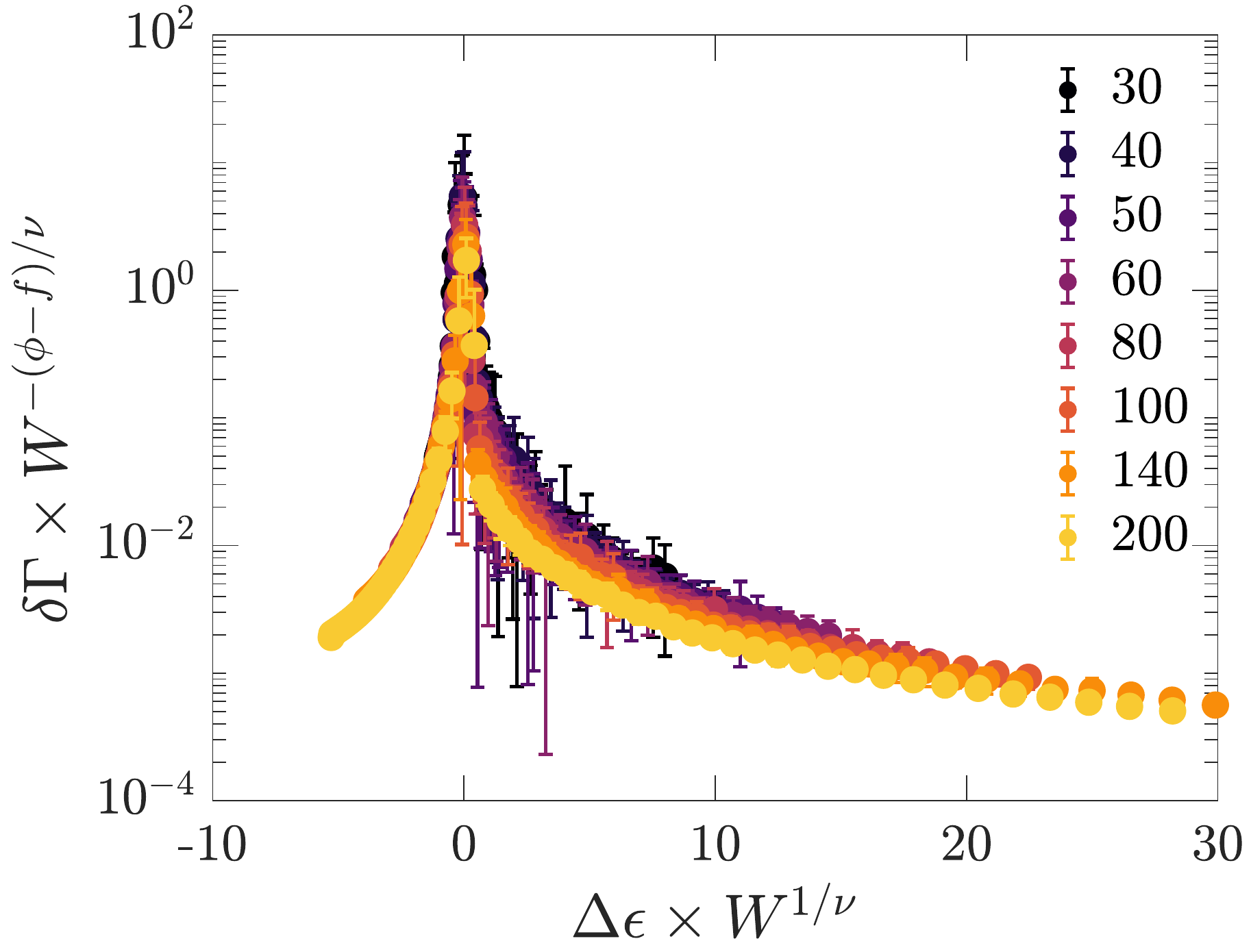}
	\caption{\label{fig_scaling_bulk_fluctuations} The finite-size scaling collapse of the nonaffine fluctuations $\delta \Gamma$ under an applied bulk strain for diluted central-force triangular networks at $z=3.3$. The correlation length exponent $\nu$ is obtained from the relation $f = d\nu - 2$.}
\end{figure}

The scaling theory in Ref.\ \cite{shivers_scaling_2019} can be generalized to athermal networks under an applied bulk strain $\varepsilon$. By following a similar renormalization procedure with $t = \varepsilon - \varepsilon_c$ and $B \sim \frac{\partial^2}{\partial t^2} h(t,\kappa)$, we derive the scaling relations in Ref.\ \cite{shivers_scaling_2019} with $f$ and $\phi$ corresponding to the scaling exponents under an applied bulk strain. In order to test the hyperscaling relation $f = d\nu -2$, we measure the nonaffine fluctuations $\delta\Gamma$ under extensional strains as
\be
\delta\Gamma = \frac{1}{Nl_c^2\delta\varepsilon^2}\sum_i{\lVert \delta \mathbf{u}_i^\mathrm{NA}\rVert^2}
\ee
in which $N$ is the number of nodes, $l_c$ is the average bond length, and $\delta\mathbf{u}_i^\mathrm{NA} = \delta\mathbf{u}_i -\delta \mathbf{u}_i^\mathrm{A}$ is the nonaffine component of the displacement of node $i$ due to the incremental strain $\delta\varepsilon$. Figure\ \ref{fig_scaling_bulk_fluctuations} shows the finite-size scaling collapse of $\delta \Gamma$ using the obtained exponents $f$ and $\phi$. The correlation length exponent $\nu$ here assumes the hyperscaling relation $\nu = (f+2)/2 \simeq 1.17$. The data collapse confirms that our scaling theory works under an applied volumetric strain.

%\FloatBarrier

\textit{Discussion} \textemdash\, In this work, we study the behavior of nonlinear bulk modulus $B$ of subisostatic fiber networks using a diluted triangular model. Similar to previous studies of fiber networks under shear \cite{sharma_strain-controlled_2016,shivers_scaling_2019}, we observe that subisostatic networks with central force interactions undergo a critical transition from a floppy to a rigid state as we increase the isotropic extensional strain. By approaching the critical extensional strain from above, we find a power law scaling behavior of $B$ with a non-mean-field exponent that is much smaller than the observed shear modulus exponent for the same model \cite{arzash_finite_2020}. By introducing finite bending rigidity $\kappa$, these networks become stable in the linear regime with $B \sim \kappa$. The stabilizing effect of $\kappa$ is evident in 
a collapse of our modulus data using a Widom-like function with two branches (Fig.\ \ref{fig_scaling_bulk_widom}). By studying the nonaffine strain fluctuations, we confirm that the recent scaling theory that was derived for fiber networks under shear also holds for a bulk deformation, although with different exponents. Our results are in agreement with the observed floppy-to-rigid transition in prior work on bulk modulus of fiber networks \cite{sheinman_nonlinear_2012}. Here, however, we focus on the critical aspects of this transition. Furthermore, based on prior studies on various fiber models in 2D and 3D \cite{licup_elastic_2016,rens_nonlinear_2016,shivers_normal_2019,arzash_stress-stabilized_2019,arzash_shear-induced_2021}, we expect to observe an analogous behavior in a different 2D or 3D system under a bulk strain.

\section*{Acknowledgments}
This work was supported in part by the National Science Foundation Division of Materials Research (Grant DMR-1826623) and the National Science Foundation Center for Theoretical Biological Physics (Grant PHY-2019745). 

%\bibliographystyle{unsrt}
%\bibliography{bulk_refs}{}


\end{document}